\documentclass[fleqn,10pt]{wlscirep}
\usepackage[utf8]{inputenc}
\usepackage[T1]{fontenc}
\usepackage{chemformula}
\let\ce\ch
\usepackage{mathtools}
\usepackage{physics}
\usepackage{floatrow}
\usepackage{hyperref}
\usepackage[justification=centering]{caption}
\usepackage{amsmath}
\usepackage{nccmath}
\usepackage{subcaption}
\usepackage{floatrow}
\usepackage[thinlines]{easytable}
\usepackage{cleveref}
\usepackage{gensymb}

\title{Radical pairs may play  a role in  microtubule reorganization}

\author[1,2,3,*]{Hadi Zadeh-Haghighi}
\author[1,2,3,*]{Christoph Simon}

\affil[1]{Department of Physics and Astronomy, University of Calgary, Calgary, AB, T2N 1N4, Canada}
\affil[2]{Institute for Quantum Science and Technology, University of Calgary, Calgary, AB, T2N 1N4, Canada}
\affil[3]{Hotchkiss Brain Institute, University of Calgary, Calgary, AB, T2N 1N4, Canada}

\affil
[*
]{hadi.zadehhaghighi@ucalgary.ca, csimo@ucalgary.ca}

\begin{abstract}
The exact mechanism behind general anesthesia remains an open question in neuroscience. It has been proposed that anesthetics selectively prevent consciousness and memory via acting on microtubules (MTs). It is known that the magnetic field modulates MT organization. A recent study shows that a radical pair model can explain the isotope effect in xenon-induced anesthesia and predicts magnetic field effects on anesthetic potency. Further, reactive oxygen species are also implicated in MT stability and anesthesia. Based on a simple radical pair mechanism model and a simple mathematical model of MT organization, we show that magnetic fields can modulate spin dynamics of naturally occurring radical pairs in MT. We show that the spin dynamics influence a rate in the reaction cycle, which translates into a change in the MT density. We can reproduce magnetic field effects on the MT concentration that have been observed. Our model also predicts additional effects at slightly higher fields. Our model further predicts that the effect of zinc on the MT density exhibits isotopic dependence. The findings of this work make a connection between microtubule based and radical pair based quantum theories of consciousness.

\end{abstract}

\begin{document}

\flushbottom
\maketitle
\thispagestyle{empty}

\section*{Introduction}

The question of how we perceive and experience the world we live in is a fascinating, long-standing open question in neuroscience, philosophy and psychology. This conscious experience vanishes during dreamless sleep or under general anesthesia \cite{koch2016neural,mashour2006integrating}. Every day millions of surgeries all over the world would not be possible without anesthetics. However, despite a century of research, the mechanisms by which anesthetics cause a reversible loss of consciousness remain obscure \cite{brown2011general}. During anesthesia, old memories are preserved, but no new memory is formed. Furthermore, anesthetics act relatively selectively on consciousness, as many non-conscious brain activities, including sensory-evoked potentials, persist during anesthesia. Understanding anesthesia, apart from its benefits in designing and developing novel anesthetics, could help explain the mystery of consciousness.\par

The connections between theories of anesthesia, related to neural correlates of consciousness, and the Meyer Overton correlation—solubility of anesthetics in a non-polar, ‘hydrophobic’ medium—has not been completely understood, and direct anesthetic effects on synaptic receptors are variable and inconsistent. Further, single-cell organisms perform cognitive activities predominantly by cytoskeletal microtubules (MTs) and are inhibited by anesthetic gases even without synapses or networks \cite{ja2015anesthetics}. Bernard showed that anesthetics act directly on cytoplasm, depending on cytoskeletal proteins' dynamics comprising actin filaments and MTs \cite{perouansky2012quest}. Franks and Lieb found that anesthetics act directly within proteins in non-polar hydrophobic regions \cite{franks1984general}. This led some anesthetic researchers to search for new target protein/molecule for anesthetics. Moreover, Eckenhoff et al. found that anesthetics bind to 23 membrane proteins and 34 cytoplasmic proteins including actin, and tubulin \cite{xi2004inhalational,pan2007halothane}. Studies suggest anesthetics exert their effects via acting on protein reaction networks involved in neuronal growth, proliferation, division, and communication, which depend on MTs \cite{pan2008inhaled}. Although the affinity of anesthetics binding to tubulin is a thousand times weaker than to membrane protein, the abundance of tubulin is thousand to ten thousand times more than membrane protein sites.\par

It has also been proposed that anesthetics act on quantum electronic activity in neuronal hydrophobic regions rather than binding to specific receptors. Consistently, Turin et al. showed that specific electron spin resonance (ESR) signals, consistent with free electrons, can be observed during anesthesia \cite{turin2014electron}. The same authors proposed that the anesthetic action may involve some form of electron transfer. Moreover, Li et al. showed experimentally that isotopes of xenon with non-zero nuclear spin had reduced anesthetic potency in mice compared with isotopes with no nuclear spin \cite{li2018nuclear}. These findings are consistent with the idea that different nuclear spins of anesthetics can modulate the electron transferring process differently. Motivated by this line of thought, a recent study shows that radical pairs may explain the mechanism behind xenon-induced anesthesia \cite{smith2021radical}.\par

Quantum physics has been proposed to be part of the solution for the mystery of consciousness. In particular the holistic character of quantum entanglement might provide an answer to the binding problem \cite{Simon2019}. In the 1990s, Penrose and Hameroff proposed a theory of consciousness based on quantum computations in MTs \cite{hameroff2014consciousness,stuart1998quantum,matsuno2001internalist,hagan2002quantum}. Computational modeling suggested that electron resonance transfer among aromatic amino acid tryptophan (Trp) rings in tubulin (subunits of MTs) in a quantum electronic process could play roles in consciousness \cite{hameroff2002conduction}. More recently, Craddock et al. showed that anesthetic molecules might bind in the same regions and hence result in anesthesia \cite{craddock2012computational}. However, quantum electronic coherence beyond ultrafast timescales demands more supporting evidence and has been recently challenged experimentally \cite{cao2020quantum}. In contrast, quantum spin coherence could be preserved for much longer timescales \cite{hu2004spin}. For example, Fisher has proposed that phosphorus nuclear spins could be entangled in networks of Posner molecules, \ch{Ca9(PO4)_6}, which could form the basis of a quantum mechanism for neural processing in the brain \cite{fisher2015quantum}. However, this particular spin-based model model also requires more supporting evidence and recently has faced experimental challenges \cite{Chen2020}. \par

It is know that magnetic fields (MFs) can influence different brain functions \cite{romero2019neural,wang2019spontaneous,lenz2016repetitive,rauvs2014extremely,pacini1999effect,grehl2015cellular,manikonda2007influence}. Recently, it has been shown that shielding the geomagnetic field—exposure to hypomagnetic field (HMF)—influences adult hippocampal neurogenesis and hippocampus-dependent cognition in mice, where ROS are implicated \cite{zhang2021long}. There exists a considerable amount of evidence showing that MFs affect MTs \cite{vassilev1982parallel,glade2005brief,bras1998susceptibility,zhang201727,qian2009large,luo2016moderate}. Wang et al. show that exposure to HMF caused tubulin assembly disorder \cite{wang2008tubulin}. Moreover, Wu et al. observe that low-frequency sub-millitesla MF modulates the density of MTs in cells \cite{wu2018weak}. All these observations establish the magnetosensitivity of MTs for wide ranges of MF strengths. \par

Magnetosensitive reactions often involve radical molecules—transient molecules with an odd number of electrons \cite{rodgers2009chemical}. A radical pair is a short-lived reaction intermediate comprising two radicals formed in non-equilibrium states whose unpaired electron spins may be in a superposition of singlet (S) and triplet (T) states \cite{Steiner1989}, depending on the parent molecule's spin configuration \cite{Timmel1998}. The radical pair mechanism (RPM) is the most promising explanation for weak magnetic field effects on chemical reactivity \cite{hore2016radical}. Schulten was the first to propose the RPM to explain the magnetoreception of migratory birds \cite{schulten1978biomagnetic}, and to date, the RPM is the most well-established model for this phenomenon \cite{xu2021magnetic}. Recently, it has also been proposed that RPM may explain xenon induced general anesthesia \cite{smith2021radical}, lithium effects on hyperactivity \cite{zadeh2021entangled} and the magnetic field effects on the circadian clock \cite{zadeh2021radical}. \par 

MTs are made of $\alpha-\beta$ tubulin heterodimers. The distribution and organization of MTs in cells are governed by a large number of MT-associated proteins (MAPs)  \cite{brouhard2018microtubule}. MTs play crucial roles in cell shape, cell transport, cell motility, cell division \cite{nogales2001structural,wu2017microtubule,akhmanova2008tracking,akhmanova2015control,redwine2012structural,monroy2020combinatorial,monroy2018competition}, neuronal polarity \cite{burute2021matrix}, information processing of living systems \cite{sanchez2021microtubules}, synaptic activity \cite{dixit2008differential}, regulating the precise timing of nerve spikes \cite{singh2021cytoskeletal}, and Alzheimer’s disease (AD) \cite{congdon2018tau}.\par

Evidence suggests that oxidative stress is vital for regulating actin and MT dynamics \cite{wilson2015regulation}. MTs contain specific amino acid residues, including Trp, tyrosine (Tyr), and phenylalanine (Phe), susceptible to oxidation. This, in turn, affects the ability of MT to polymerize and causes the severing of actin microfilaments in neuronal and non-neuronal cells. Contrarily, ROS inhibition causes aberrations in actin polymerization, decreases neurite outgrowth, and affects neurons' normal development and polarization. \par

Various studies have proposed mathematical models for the dynamics and stability of MTs \cite{white2015microtubule,bowne2013microtubule}. Craddock et al. show that the dynamics of MT can be framed in a simple kinetic model \cite{craddock2012zinc}. In the context of the RPM, Player et al. show that quantum effects can be introduced to the chemical oscillator model by considering the quantum effects on the corresponding reaction rates in the chemical equations \cite{player2021amplification}. Taking the same approach, a new study shows that quantum effects can directly modulate the period of the circadian clock in \textit{Drosophila}, where the spin dynamics of the PRs are the key elements \cite{zadeh2021radical}.\par

\begin{figure}[ht]
  \includegraphics[width=0.55\linewidth]{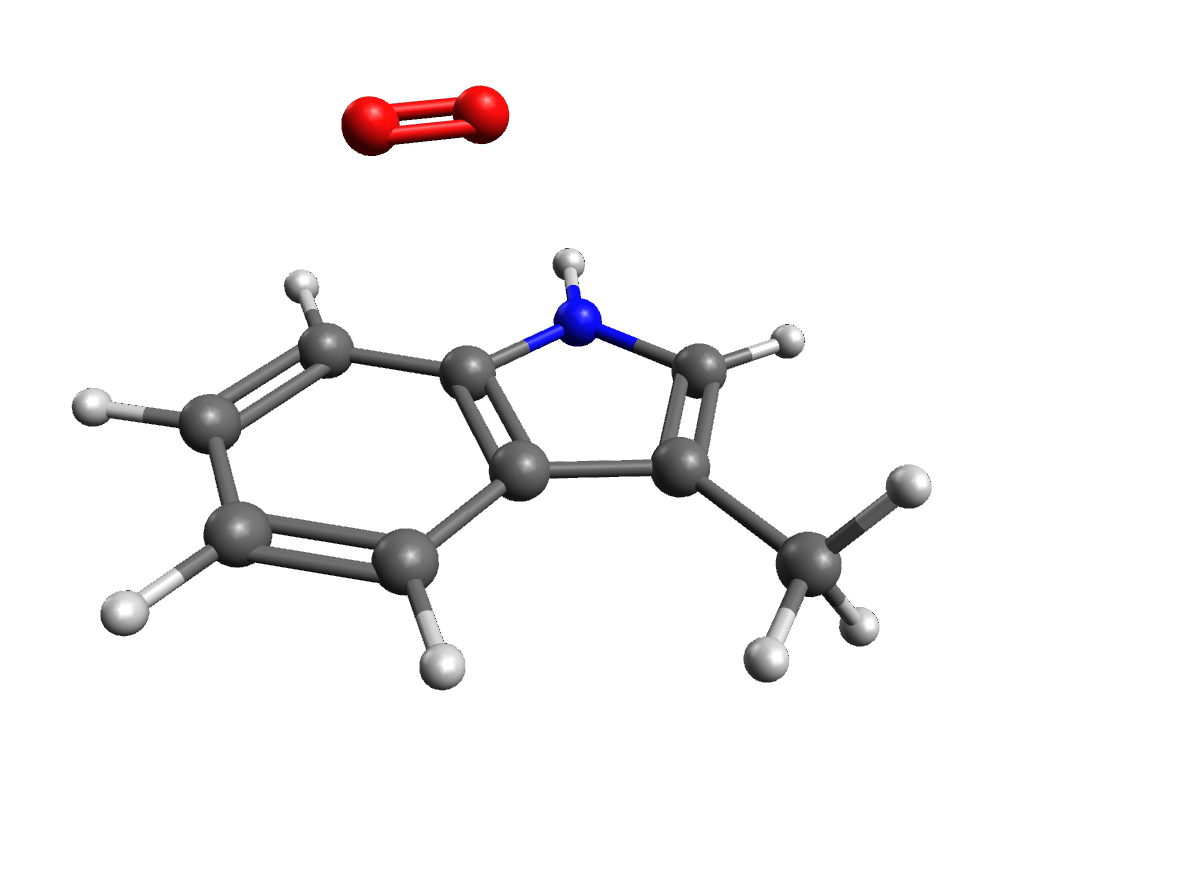}
  \caption{Schematic presentation of \ch{TrpH^{.+}} [the aromatic molecule] and \ch{O2^{.-}} [the red molecule] radical pair, considered in the RPM model in the present work, similar to Ref. \cite{smith2021radical}. The radical pair undergoes interconversion between singlet and triplet states. Image rendered using Avogadro (https://avogadro.cc).} 
\label{fig:schem}
\end{figure}

In the present work, we propose that there are naturally occurring RPs in the form of \ch{TrpH^{.+}} and \ch{O2^{.-}}, which is an important ROS, that play important roles in MT organization, as shown in Fig. \ref{fig:schem}. Our model predicts that the applied magnetic fields alter the spin dynamics of such RPs, similar to Ref. \cite{smith2021radical}, and hence modulate the assembly of MTs in cytoskeleton. It
further predicts that the effect of zinc on the MT density exhibits isotopic dependence.\par

In the following, we review the experimental results for the effects of applied magnetic field \cite{wu2018weak,wang2008tubulin} on the density of MTs. We then describe the quantum spin dynamics of our radical pair model by a simple kinetic model for the MT dynamics. Next, we introduce the quantum effect to the density of the MT, and we show that our model can reproduce the observed magnetic field effects; it further makes new predictions for experiments using low-frequency MFs. Lastly, the model predicts that the zinc effect on the MT density is isotope dependent.  

\section*{Results}
\subsection*{Prior experiment}
Wang et al. report that shielding Earth's geomagnetic field (GMF) (0.025-0.065 mT) caused disorders in tubulin self-assembly \cite{wang2008tubulin}. They show that the absorbance at 350 nm, which is for monitoring tubulin self-assembly, was altered by exposure to 30 min HMF. Average gray volume per cell was reported based on the amount of fluorescence in each cell. About 95$\%$ of tubulin were assembled in the GMF, while much less of the tubulin assembled ($\sim$ 64$\%$) in the HMF. In that work, the magnitude of the residual GMF was 10–100 nT.

\subsection*{Radical pair model calculations}
We develop an RP model to reproduce the HMF effects on the MT density observed by Wang et al. \cite{wang2008tubulin}. Among the amino acids in MTs, as mentioned above, Trp is redox active \cite{saito1981formation}, as shown by its involvement in RP formation in the context of cryptochrome, and could feasibly participate in the creation of RPs. We propose that the magnetic field interacts with the spins of RPs on Trp and superoxide. Note, that superoxide is thought to form RPs with flavins in other contexts \cite{usselman2016quantum}. The correlated spins of RP are assumed to be in the [\ch{TrpH^{.+}} ... \ch{O2^{.-}}] form, following Ref. \cite{smith2021radical}, where the unpaired electron on each molecule couples to the nuclear spins in the corresponding molecule. Oxygen has zero nuclear spin and thus zero coupling with its nucleus.\par

We consider a simplified system in which the unpaired electron on \ch{TrpH^{.+}} is coupled to the Trp's $\beta$-proton with the largest isotropic HF coupling constant (HFCC) of 1.6046 mT \cite{lee2014alternative} among all the nuclei in Trp. We consider only Zeeman and HF interactions \cite{Efimova2008,hore2016radical}. For the RPs, we assume the $g$-values of a free electron (which is an excellent approximation in the low-field regime that we are considering here). The Hamiltonian for the RP system reads:    

\begin{ceqn}
\begin{equation}
    \hat{H}=\omega \hat{S}_{A_{z}}+a_A \mathbf{\hat{S}}_A.\mathbf{\hat{I}}_1+\omega \hat{S}_{B_{z}},
\end{equation}
\end{ceqn}
where $\mathbf{\hat{S}}_A$ and $\mathbf{\hat{S}}_B$ are the spin operators of radical electron on \ch{TrpH^{.+}} and \ch{O2^{.}}, respectively, $\mathbf{\hat{I}}_A$ is the nuclear spin operator of \ch{TrpH^{.+}}'s $\beta$-proton, $a_{A}$ is HFCC, and $\omega$ is the Larmor precession frequency of the electrons due to the Zeeman effect. In the model presented here, for zinc effects, $a_{B}$ corresponds to the nuclear spin of zinc. We assumed the initial state of RPs to be singlet states (see the Discussion section).\par

Using the Liouville-von Neumann equation for the spin state of the radical pair, we calculate the singlet yield resulting from the radical pair mechanism throughout the reaction. The ultimate triplet yield, $\Phi_T$, for periods much greater than the RP lifetime \cite{Hore2019} has the following form:

\begin{ceqn}
\begin{equation}
    \Phi_T=\frac{3}{4}+\frac{k}{4(k+r)}-\frac{1}{M}\sum_{m=1}^{4M}\sum_{n=1}^{4M}|\bra{m}\hat{P}^S \ket{n}|^2 \frac{ k(k+r)}{(k+r)^2+(\omega_m-\omega_n)^2},
    \label{eq:sy}
\end{equation}
\end{ceqn}
where $M$ is the nuclear spin multiplicity, $\hat{P}^S$ is the singlet projection operator, $\ket{m}$ and $\ket{n}$ are eigenstates of $\hat{H}$ with corresponding eigenenergies of $\omega_m$ and $\omega_n$, respectively, $k$ is the RP reaction rate, and $r$ is the RP spin-coherence rate (relaxation rate). In this model, we assumed the reaction rates for singlet and triplet have the same values. Of note, the singlet and triplet product of the RP system in [\ch{TrpH^{.+}} ... \ch{O2^{.-}}] are \ch{H2O2} and \ch{O2^{-}} \cite{usselman2016quantum}, respectively, which are the major ROS in redox regulation of biological activities and signaling \cite{sies2020reactive}.\par

\begin{figure}[H]
  \includegraphics[width=0.6\linewidth]{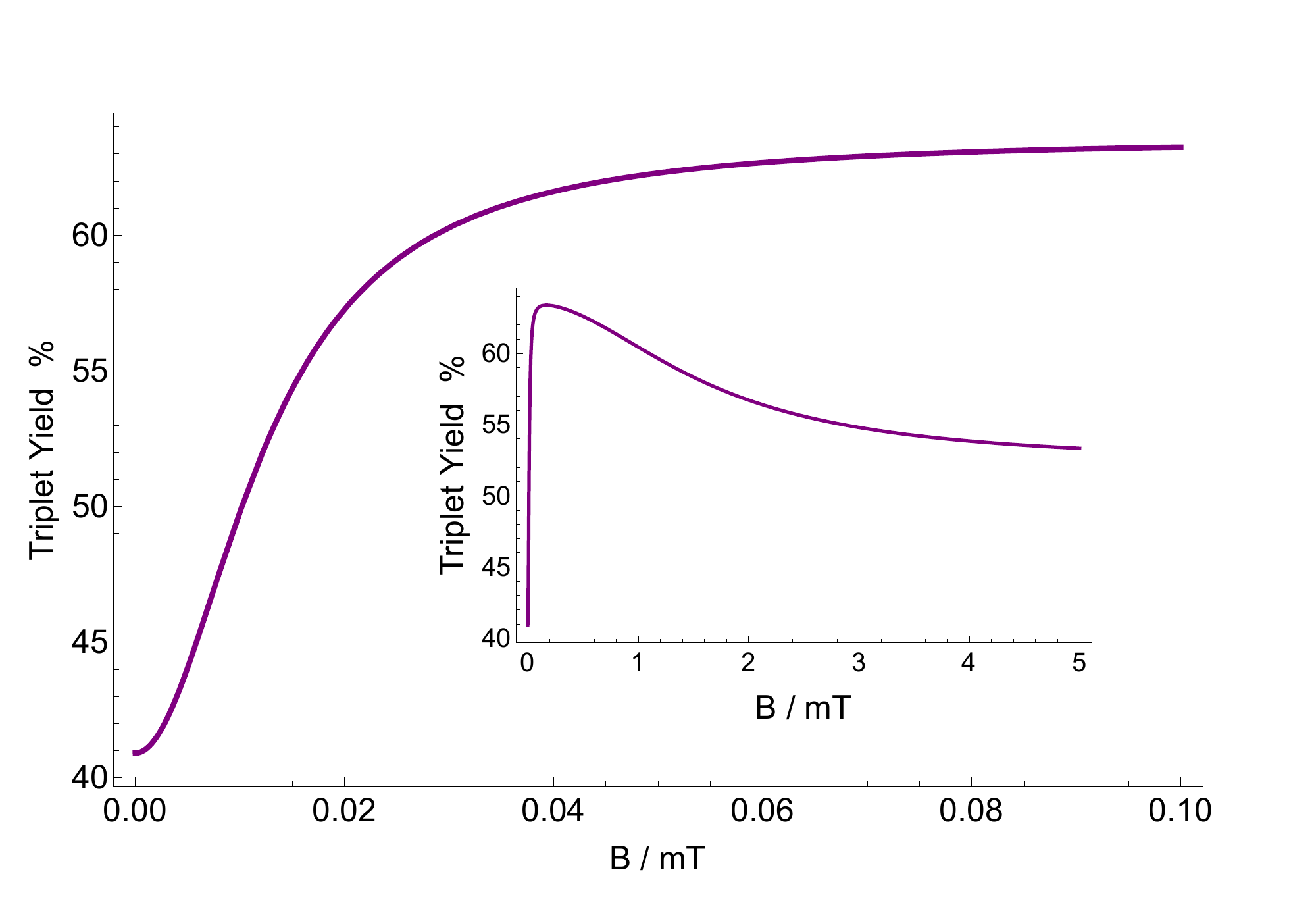}
  \caption{The dependence of the triplet yield of the [\ch{TrpH^{.+}} ... \ch{O2^{.-}}] complex on applied static magnetic field for $r=10^{5}$ s$^{-1}$, $k=10^{6}$ s$^{-1}$, $a_{A}=1.6046$ mT. The triplet yield goes from around 60\% to around 40\% by changing the magnetic field strength from GMF to HMF. The strong HMF effect is further emphasized by the inset.} 
\label{fig:TY-SB}
\end{figure}

Here we look at the dependence of the triplet yield on changes in the strength of the external static magnetic field for the [\ch{TrpH^{.+}} ... \ch{O2^{.-}}] radical complex, as shown in Fig. \ref{fig:TY-SB}, for $k=10^{6}$ s$^{-1}$, and $r=10^{5}$ s$^{-1}$ with $a_{A}=1.6046$ mT. One can notice a fairly strong HMF effect (the triplet yield goes from around 60\% to around 40\%). Our choices for the rate constants are discussed below and in the Discussion section. \par

\subsection*{Chemical kinetics model for MT dynamics}
We use a simple mathematical model for the dynamics of MT, following the work of Craddock et al. \cite{craddock2012zinc}. This model is based on the interconversion of the free tubulins, Tu, and the MT in cytoskeleton:

\begin{ceqn}
\begin{equation}\label{eq:chem}
\ch{Tu <>[ $k_{\mathrm{p}}$ ][ $k_{\mathrm{d}}$ ] MT},
\end{equation}
\end{ceqn}
where $k_p$ and $k_d$ are the polymerization rate and the depolymerization rate, respectively. The chemical equation reads as follows:

\begin{ceqn}
\begin{equation}
    \frac{[dMT(t)]}{dt}=k_p [Tu(t)]-k_d [MT(t)],
    \label{eq:mt}
\end{equation}
\end{ceqn}
which yields $[MT(t)]= \frac{k_p [P]}{k_p+k_d} (1-e^{-(k_p+k_d)t})$.

Craddock et al. suggest that changes in the concentration of zinc can be incorporated by modulating the polymerization rate \cite{craddock2012zinc}. In the present work, it is assumed that the total tubulin protein concentration [P] is 400 $\mu$M, such that [MT]+[Tu]=[P], $k_p$ = 90 s$^{-1}$ and $k_d$ = 150 s$^{-1}$. These values are chosen such that we could be able to reproduce the experimental findings, which are comparable to those in Ref. \cite{craddock2012zinc}.

\subsection*{Quantum effects on microtubule density}
The effects of triplet yield change can be introduced to the chemical equation of the MTs by modifying the kinetic rates, following the work of Player et al.  \cite{player2021amplification,zadeh2021radical}. In the chemical kinetics equations for MT oscillations, Eq. \ref{eq:mt}, the corresponding rate is $k_{p}$, which is the MT polymerization rate \cite{craddock2012zinc}. The key assumption in our model is that this rate is influenced by a RP reaction. The effect of the triplet yield change on $k_{p}$ reads:

\begin{ceqn}
\begin{equation}\label{eq:yeild-chem-n}
k'_{p} \propto k_{p} \frac{\Phi_T'}{\Phi_T},
\end{equation}
\end{ceqn} 
where $k'_{p}$, $\Phi_T$, and $\Phi'_T$ are the modified rate constant $k_{p}$, the triplet yield under natural quantum effects (only GMF and no isotope effects), and the triplet yield resulting from quantum effects due to the external MF effects and hyperfine interactions from isotope effects, respectively.

\subsubsection*{Magnetic field effects on microtubule density}
\begin{figure}[H] 
  \includegraphics[width=0.6\linewidth]{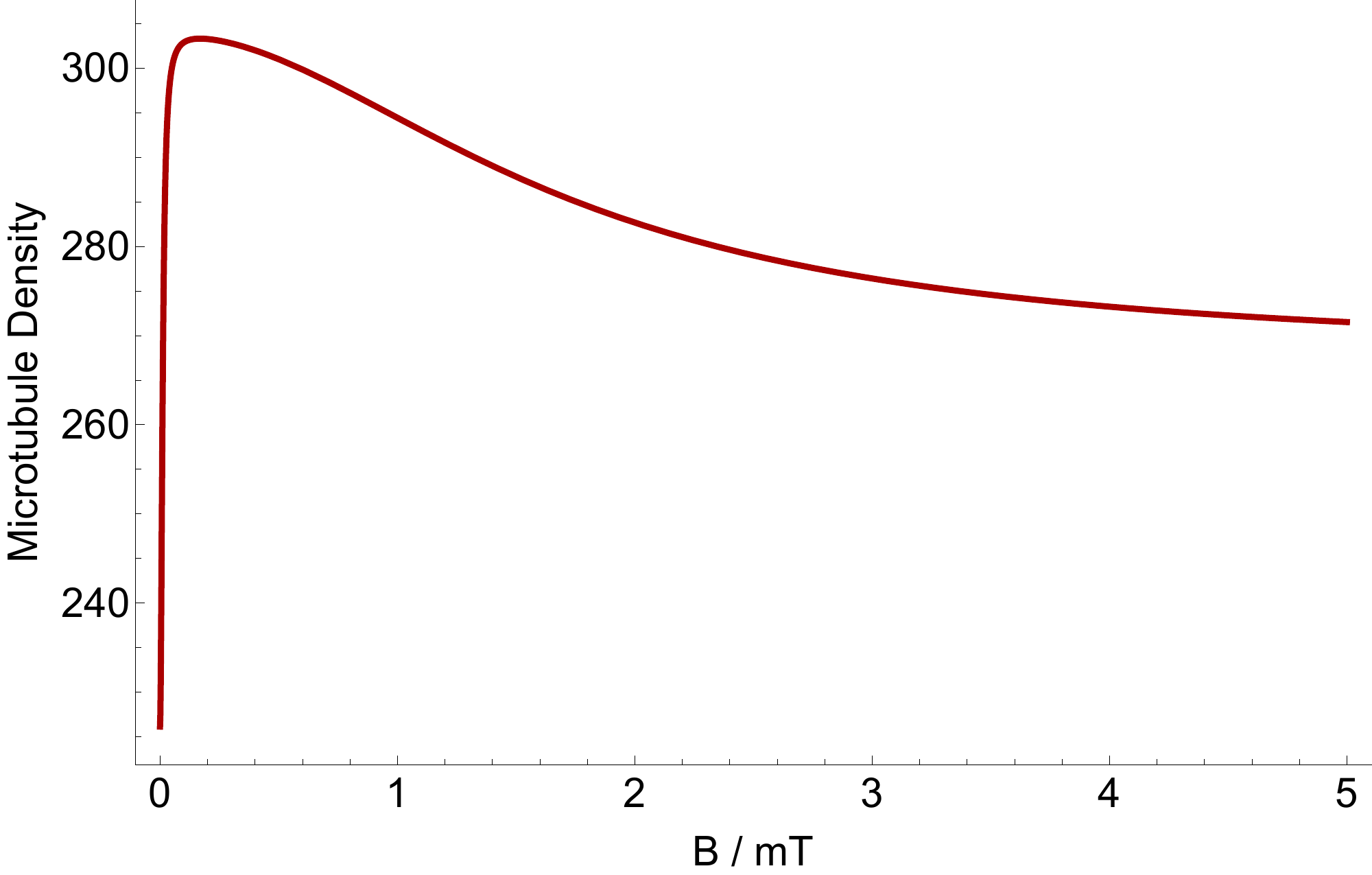}
  \caption{The dependence of the microtubule density ratio in geomagnetic field compared to applied static magnetic field based on the RP complex of [\ch{TrpH^{.+}} ... \ch{O2^{.-}}]. The geomagnetic field is 0.05 mT. The magnetic field modulates rate $k_p$ for $a_{A}=1.6046$ mT, $r=10^{5}$ s$^{-1}$, and $k=10^{6}$ s$^{-1}$. HMF causes strong decrease on the microtubule density. The maximum microtubule density occurs around GMF.} 
\label{fig:mt-sb}
\end{figure}

Here, we look at the explicit effects of an applied magnetic field on the density of microtubules. Using Eqs. \ref{eq:mt}, we explore the parameter space of relaxation rate $r$ and recombination rate $k$ in order to investigate the effects of shielding geomagnetic field on MT's density. Wang et al. report that the ratio MT density of geomagnetic field over hypomagnetic field is about 1.48 \cite{wang2008tubulin}. Fig. \ref{fig:contT} show that the ratio of the MT density of GMF over HMF can reach above 1.3, which has the right trend compared to the experimental findings. However the uncertainty of the experiment was not reported. Our model predicts a magnetic dependence of the MT density. Fig. \ref{fig:mt-sb} show the dependence of the MT density ratio of in GMF compared to applied static magnetic field, for $a_{A}=1.6046$ mT based the RP complex of [\ch{TrpH^{.+}} ... \ch{O2^{.-}}], $k=10^{6}$ s$^{-1}$, and  $r=10^{5}$ s$^{-1}$. Fig. \ref{fig:mt-sb} indicates that exposure to static magnetic fields stronger than the geomagnetic field could decrease the the microtubule density ratio. The maximum MT density occurs around 0.05 mT, which is in the range of the GMF.

\subsubsection*{Zinc isotope effect on microtubule density}
The RPM typically leads to isotope effects \cite{smith2021radical,zadeh2021entangled}, and thus an isotope effect would be a good test of our proposal. It is known that alterations in zinc ion concentration in neurons influence the stability of polymerized MTs \cite{craddock2012zinc}. Zn is a positively charged ion. Thus it is natural to assume that \ch{Zn^{2+}} couples with a molecule with a negative charge. Among all stable isotopes of zinc, only \ce{^{67}Zn} has a nuclear spin of $I_{B}=-\frac{5}{2}$, with a natural abundance of 4$\%$. Here we explore the isotope effect of zinc on the density of MTs. In this model nuclear spin of zinc modulates the spin dynamics of the RPs via hyperfine interactions. The Hamiltonian for the RP system here reads:
\begin{ceqn}
\begin{equation}
    \hat{H}=\omega \hat{S}_{A_{z}}+a_A \mathbf{\hat{S}}_A.\mathbf{\hat{I}}_1+\omega \hat{S}_{B_{z}}+a_B \mathbf{\hat{S}}_B.\mathbf{\hat{I}}_2,
    \label{eq:zn-ham}
\end{equation}
\end{ceqn}
where $\mathbf{\hat{I}}_2$ is the nuclear spin operator of \ch{Zn^{+2}}, and $a_B=-11.39$ mT is the HFCC of \ch{Zn^{+2}}.\par

\begin{figure}[H] 
  \includegraphics[width=0.6\linewidth]{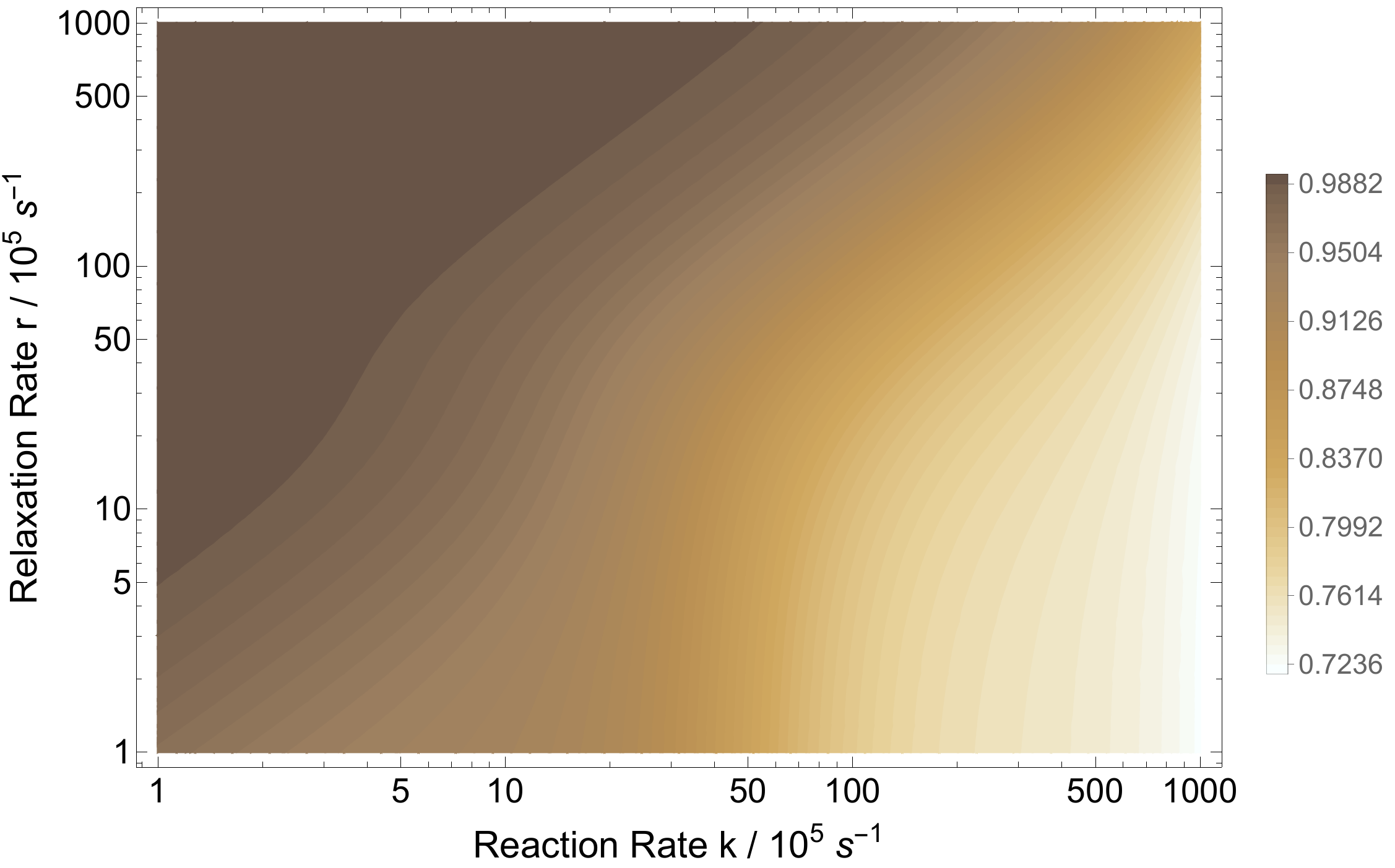}
  \caption{The RPM model prediction of the MT density ratio of the administration of Zn (with zero nuclear spin) over \ce{^{67}Zn} (with nuclear spin of $I_{B}=-\frac{5}{2}$) based on the RP complex of [\ch{TrpH^{.+}} ... \ch{O2^{.-}}]. \ce{^{67}Zn} The geomagnetic field has a strength of 0.05 mT. \ce{^{67}Zn}'s nuclear spin modulates rate $k_p$ for $a_{A}=1.6046$ mT and $a_{B}=-11.39$ mT.} 
\label{fig:cont-zn}
\end{figure}

\begin{figure}[H] 
  \includegraphics[width=0.6\linewidth]{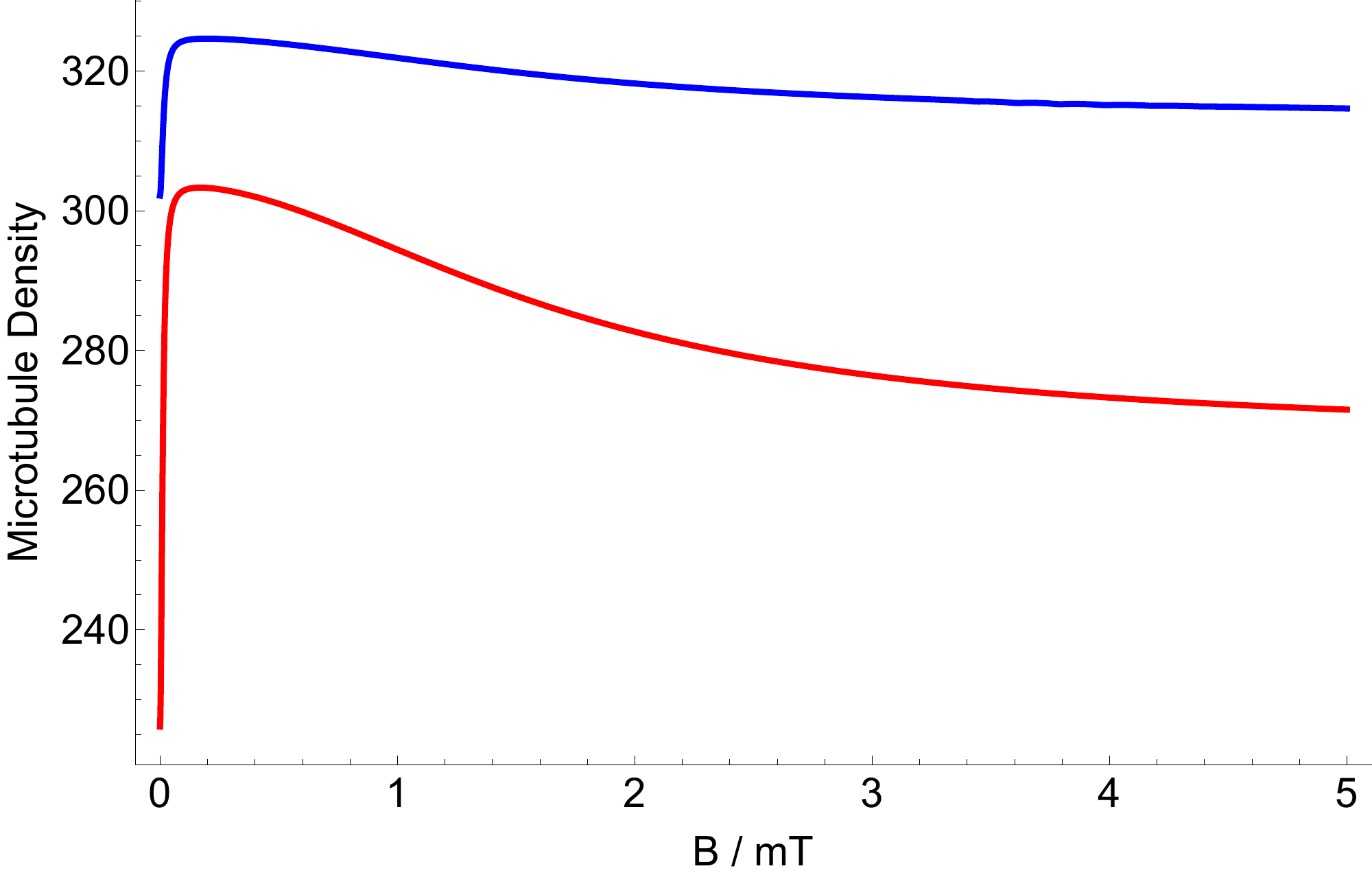}
  \caption{The dependence of the MT density of the administration of Zn (with zero nuclear spin) [Red] and \ce{^{67}Zn} (with nuclear spin of $I_{B}=-\frac{5}{2}$) [Blue] on the strength of applied magnetic field based on the RP complex of [\ch{TrpH^{.+}} ... \ch{O2^{.-}}]. The geomagnetic field is 0.05 mT. The magnetic field modulates rate $k_p$ for $a_{A}=1.6046$ mT, $a_{B}=-11.39$ mT, $r=10^{5}$ s$^{-1}$, and $k=10^{6}$ s$^{-1}$.} 
\label{fig:mt-sb-zn}
\end{figure}

We look at the effect of \ce{^{67}Zn} on the MT density. We assume that \ce{^{67}Zn} interacts with the superoxide radical. Our model predicts that administration of \ce{^{67}Zn} increases the density of MTs compared to Zn with zero nuclear spin, as shown in Fig.\ref{fig:cont-zn}. We explored the parameter space of relaxation rate $r$ and recombination rate $k$ in order to investigate the effects of the \ce{^{67}Zn} treatment on MT's density, for $a_{A}=1.6046$ mT and $a_{B}=-11.39$ mT, as shown in Fig. \ref{fig:cont-zn}. Fig. \ref{fig:mt-sb-zn} shows the dependence of the MT density ratio of the administration of Zn over \ce{^{67}Zn} on the strength of applied magnetic field based on the RP complex of [\ch{TrpH^{.+}} ... \ch{O2^{.-}}]. The geomagnetic field is 0.05 mT. The magnetic field modulates rates $k_d$ and $k_p$ for $r=10^{5}$ s$^{-1}$ and $k=10^{6}$ s$^{-1}$. Our model predicts that administering \ce{^{67}Zn} increases the MT density compared to Zn without nuclear spin.

\section*{Discussion}
\begin{figure}[H] 
  \includegraphics[width=0.6\linewidth]{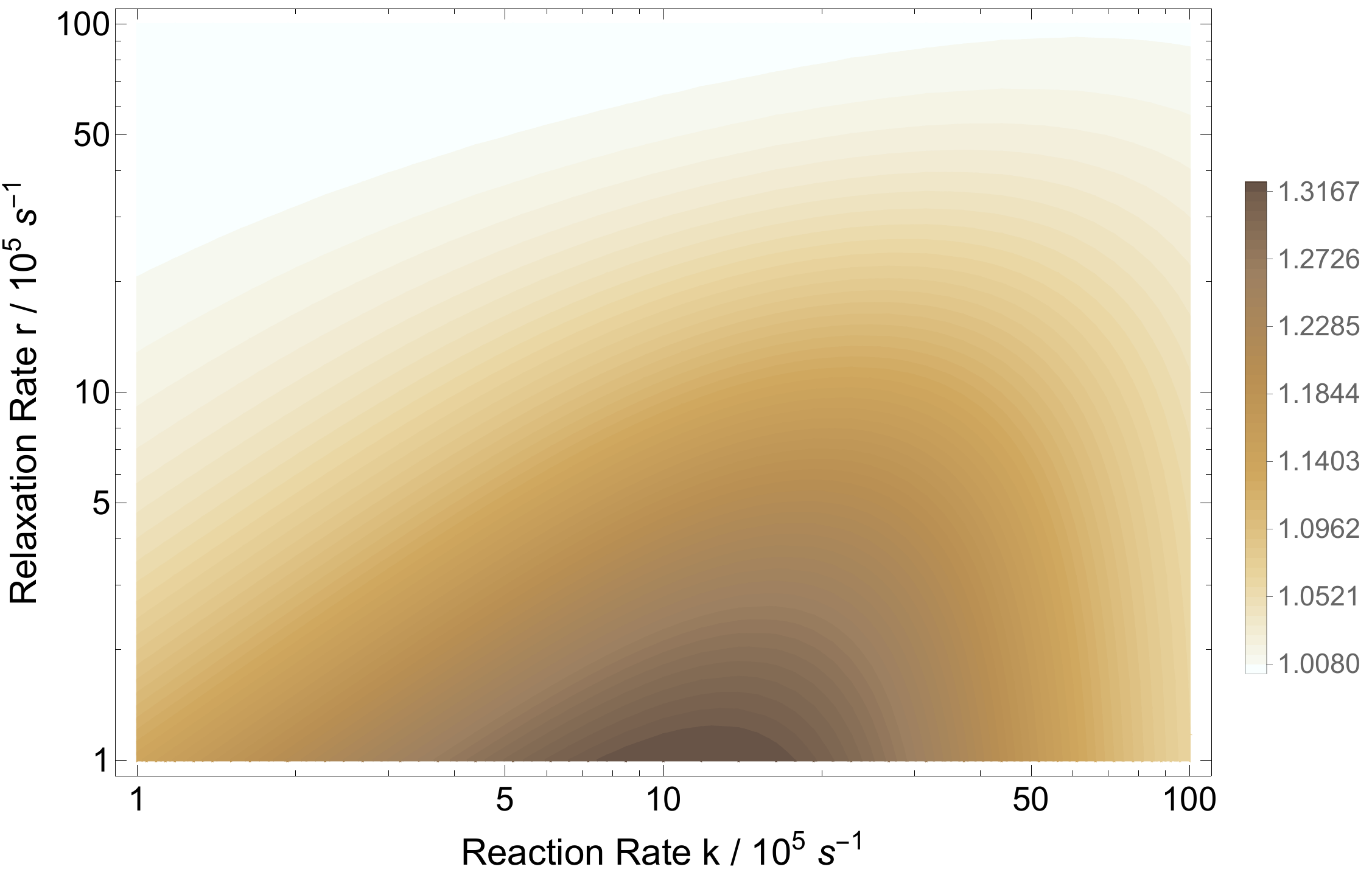}
  \caption{The RPM model prediction of the microtubule density ratio in geomagnetic field compared to hypomagnetic field based on the RP complex of [\ch{TrpH^{.+}} ... \ch{O2^{.-}}]. The geomagnetic field and hypomagnetic field are 0.05 mT and 10 nT, respectively. The magnetic field modulates rate $k_p$ for $a_{A}=1.6046$ mT. The maximum HMF effect is obtained for $k \in [7\times10^5,11\times10^5]$ s$^{-1}$ and $r \in [10^5,2\times10^5]$ s$^{-1}$.} 
\label{fig:contT}
\end{figure}

In the present work, our main goal was to explore whether an RP model can help explain the magnetic field effects on the MT density. We showed that the quantum effects influence the rates in the chemical kinetics of the MT dynamics, and then this results in a change in the density of MT in cytoskeleton. Our model reproduces the experimental findings of Ref. \cite{wang2008tubulin} fairly well, as shown in Fig. \ref{fig:contT}. Our model further predicts that exposure to static magnetic fields stronger than the geomagnetic field could decrease the the microtubule density ratio, with the maximum microtubule density occurs at a magnetic field strength in the range of the geomagnetic field, which might have evolutionary roots.\par

The present work predicts that the zinc effects on the MT density exhibits an isotope dependent manner. Isotope effects are generally a good indication for radical pairs. It thus would be interesting to perform an experiment to probe such isotope effects.\par

The \ch{O2^{.-}} radical is thought to have a fast spin relaxation rate \cite{Player2019,Hogben2009}. However, it has also been proposed that this fast spin relaxation can be decreased by the biological environment. Additionally, Kattnig and Hore show that scavenger species around the superoxide radical can also reduce its fast spin relaxation \cite{Kattnig2017,Kattnig2017b}. \par 

The ground state of the oxygen molecule is singlet, and due to this, it is often assumed that in the RP complexes superoxide are formed in triplet states, as opposed to the case considered in the present work, where it is assumed to be formed as a singlet. Studies have shown the formation of superoxide from singlet oxygen \cite{saito1981formation}. Thus it seems reasonable that the initial state for RP formation could also be its excited singlet state, which is also which is its excited state (and is also a biologically relevant ROS) \cite{kerver1997situ,miyamoto2014singlet,kanofsky1989singlet}. Further, the transition of the initial RP state from triplet to singlet could also take place due to spin-orbit coupling \cite{goushi2012organic,fay2019radical}. \par

Let us also note that this model could be adapted for other RP complexes in the microtubule dynamics, e.g. involving flavin and tyr. Such radical pairs can be formed via flavoenzymes \cite{liu2016mical3,yamauchi1983disassembly}.\par

A well-known indication of the pathogenesis of tauopathy—loss-of-function effects on the microtubules and the gain-of-function effects of the toxic tau species—is oxidative stress, which contributes to tau phosphorylation and the formation of neurofibrillary tangles \cite{ballatore2007tau}. However, the mechanisms behind the connection between reactive oxygen species (ROS) generated by oxidative stress and tau hyperphosphorylation are elusive \cite{kurian2017oxidative}. Further, redox signaling and oxidative stress regulate cytoskeletal dynamics \cite{wilson2015regulation}. Proper balances between chemical reduction and oxidation (known as redox balance) are crucial for normal cellular physiology. Imbalances in the production of oxidative species lead to DNA damage, lipid peroxidation, and aberrant post-translational modification of proteins, which could induce injury, cell death, and disease \cite{sies2020reactive}. These findings further emphasize the role of radicals and likely the need for a RPM in the context of the brain. Our proposed RPM model for magnetic field effects and Zn isotope effects on the microtubule organization includes superoxide radical.\par

Microtubules not only play crucial roles in cell shape, cell transport, cell motility, and cell division, but also are important targets for curing devastating diseases, such as Alzheimer's disease \cite{gong2008hyperphosphorylation,alonso1996alzheimer,iqbal2010tau}, Parkinson's diseases \cite{feng2006microtubule,kett2012lrrk2}, and cancer \cite{parker2014microtubules,mukhtar2014targeting,jordan2004microtubules}. The dynamics of MAPs signaling proteins play critical roles in the MT network, hence in the process of synaptic plasticity and brain function. Studies suggest that memory is encoded in MT of neuronal dendrites and cell bodies. Anesthetic causes loss of consciousness and memory formation via acting on MTs \cite{craddock2012cytoskeletal,janke2010tubulin,janke2011post}. Disintegration and separation of the MT-associated protein tau have been observed in memory neurodegenerative diseases and disorders, e.g., in AD \cite{goldstein2012chronic}; MT-stabilizers is currently the target for during such diseases \cite{brunden2010epothilone,congdon2018tau,li2017tau}. However, the underlying mechanism for such diseases is mostly unknown.Thus this project also paves a potential path to study other functionalities of the body and the brain connected to MTs in the light of the RPM. \par

In conclusion, our results suggest that quantum effects may underlie the magnetic field effects on the microtubule dynamics. This is likely a similar mechanism to those behind magnetoreception in animals \cite{xu2021magnetic}, xenon-induced general anesthesia \cite{smith2021radical}, lithium treatment for mania \cite{zadeh2021entangled}, the magnetic field effects on the circadian clock \cite{zadeh2021radical}. Our work is thus another piece of evidence that quantum entanglement \cite{fisher2015quantum,Simon2019,Adams2020,gauger2011sustained,bandyopadhyay2012quantum,cai2010quantum,kominis2012magnetic,pauls2013quantum,tiersch2014approaches,zhang2014sensitivity} may play essential roles in the brain's functions, anesthesia, and consciousness \cite{Hameroff2014b,hameroff2014consciousness}. Particularly, the photo-emission of singlet oxygen could serve as quantum messengers to establish long-distance connections \cite{Kumar2016} that might be essential for consciousness. Our work also provides a potential connection between microtubule-based and spin-based quantum approaches to consciousness.

\section*{Methods} 
\subsection*{DFT Analysis}
\begin{figure}[H] 
  \includegraphics[width=0.36\linewidth]{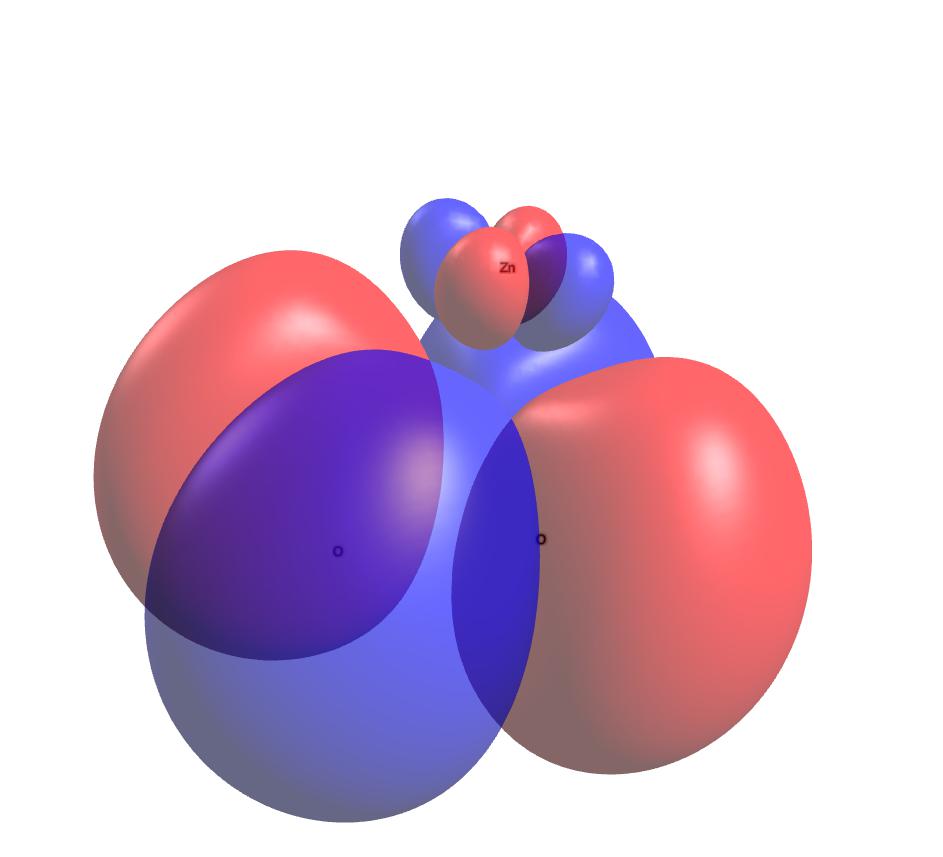}
  \caption{The highest occupied molecular orbital of [\ch{Zn^{2+}}-\ch{O2^{.-}}]. Imaged rendered using Avogadro (https://avogadro.cc).} 
\label{fig:homo}
\end{figure}

The ORCA package \cite{Neese2011} was used for our \ch{Zn^{2+}}-\ch{O2^{.-}} DFT calculations, and the molecular structure was optimized using PBE0/def2-TZVP. The orbitals obtained from the optimization calculations were used to calculate orbital energies as well as the hyperfine coupling constant $a_B$. Using RI-B2GP-PLYP/def2-QZVPP\cite{Goerigk2011}, we obtained $a_2=-11.39$ mT. In these calculations, relativistic effects were treated by a scalar relativistic Hamiltonian using the zeroth-order regular approximation (ZORA) \cite{vanLenthe1996}. Solvent effects were considered by using the conductor-like polarizable continuum model (CPCM) \cite{Marenich2009}, with a dielectric constant of 2. The resulting Mulliken charge and spin population of the [\ch{Zn^{2+}}-\ch{O2^{.-}}] complex indicates that the unpaired electron resides primarily on the \ch{O2} molecule but is extended slightly onto the zinc atom, see Table \ref{tab:mulliken}. The highest occupied molecular orbital (HOMO) of [\ch{Zn^{2+}}-\ch{O2^{.-}}] is shown in Fig. \ref{fig:homo}.\par

\begin{table}[tbhp]
\centering
\caption{\label{tab:mulliken}Mulliken charge and spin population of [\ch{Zn^{2+}}... \ch{O2^{.-}}].} 
\begin{tabular}{ccc}
Atom & Charge Population &Spin Population\\
\hline
O & -0.203235 & 0.554431 \\

O & -0.203243 &   0.554413 \\

Zn &  1.406478  & -0.108844 \\
\hline
Sum & 	1 & 1 \\
\end{tabular}
\end{table}

\section*{Data Availability}
The generated datasets and computational analysis are available from the corresponding author on reasonable request.

\bibliography{sample}

\section*{Acknowledgements}
The authors would like to thank Jack Tuszyński for valuable discussions. This work was supported by the Natural Sciences and Engineering Research Council of Canada.

\section*{Author contributions statement}
H.ZH. and C.S. conceived the project; H.ZH. performed the calculations; H.ZH. and C.S. wrote the paper; C.S. supervised the project.

\section*{Competing Interests}
The authors declare no competing interests.

\end{document}